\documentclass{aa}
\usepackage{graphicx}
\usepackage{txfonts}
\usepackage{natbib}
\bibpunct{(}{)}{;}{a}{}{,}
\begin{document}
\title{High-resolution 21-cm observations of low-column density 
gas clumps in the Milky Way halo} 
\titlerunning{Low-column density gas clumps in the halo of the Milky Way}
\subtitle{}
\author{N. Ben Bekhti\inst{1}  \and  P. Richter\inst{2} \and B. Winkel\inst{1} 
\and F. Kenn\inst{1} \and T. Westmeier\inst{3} }
\offprints{N. Ben Bekhti}
\institute{Argelander-Institut f\"ur Astronomie, Universit\"{a}t Bonn, Auf 
dem H\"{u}gel 71, 53121 Bonn, Germany\\
\email{nbekhti@astro.uni-bonn.de},
\email{fkenn@astro.uni-bonn.de}
\email{bwinkel@astro.uni-bonn.de}\and Institut f\"{u}r Physik und Astronomie, 
Haus 28, Karl-Liebknecht-Str. 24/25, 14476 Golm (Potsdam), Germany\\
\email{prichter@astro.physik.uni-potsdam.de} \and Australia Telescope
National Facility, PO Box 76, Epping NSW 1710, Australia\\ 
\email{tobias.westmeier@csiro.au}}
\date{Received month ??, ????; accepted month ??, ????}

\abstract{}{We study the properties of low-column density gas clumps 
in the halo of the Milky Way based on high-resolution 21-cm observations.}
{Using interferometric data from the Westerbork Synthesis Radio Telescope (WSRT) 
and the Very Large Array (VLA) we study \ion{H}{i} emission at low-, intermediate- and 
high radial velocities along four lines of sight towards the quasars QSO\,J0003$-
$2323, QSO\,B1331$+$170, QSO\,B0450$-$1310, and J081331$+$254503. Along these 
sightlines we previously have detected weak     
\ion{Ca}{ii} and \ion{Na}{i} absorbers in the optical spectra of these quasars.}
{The analysis of the high-resolution \ion{H}{i} data reveals the presence of 
several compact and cold clumps of neutral gas at velocities similar to the 
optical absorption. The clumps have narrow \ion{H}{i} line widths in the 
range of $1.8 \leq v_\mathrm{FWHM} \leq 13.0$\,km\,s$^{-1}$, yielding upper 
limits for the kinetic temperature of the gas of $70 \leq T_\mathrm{max} \leq 
3700$\,K. The neutral gas has low \ion{H}{i} column densities in the range of 
$5 \cdot 10^{18} \ldots3 \cdot 10^{19}$\,cm$^{-2}$. All clumps 
have angular sizes of only a few arcminutes.}
{Our high-resolution 21-cm observations indicate that many of the  \ion{Ca}{ii} and \ion{Na}{i} absorbers seen in our optical quasar spectra 
are associated with low-column density \ion{H}{i} clumps 
at small angular scales. This suggests that next to the massive, high-column 
density neutral gas clouds in the halo (the common 21-cm low-, intermediate-, and high-velocity clouds, LVCs, IVCs, and HVCs) there exists a population of low-mass, neutral 
gas structures in the halo that remain mostly unseen in the existing 21-cm all-sky surveys of IVCs and HVCs. One of our absorbers may be associated with the
Magellanic Stream, two intermediate-velocity clouds are probably part of the Intermediate-Velocity Spur and the Low-latitude IV arch, respectively. The remaining systems could be located either in the lower halo or in the disk of the Milky Way. The estimated thermal gas 
pressures of the detected \ion{H}{i} clumps are consistent with what is expected 
from theoretical models of gas in the inner and 
outer Milky Way halo.}


\maketitle

\section{Introduction}

The disk of the Milky Way is surrounded by a gaseous halo that has an unknown 
extent. High-resolution absorption and emission line measurements have 
demonstrated that this halo contains a mixture of cool, warm and hot gas 
\citep[e.g.,][ and references therein]{savage_massa87, majewski04, 
fraternalietal_07}. Multi-wavelength observations of this gaseous material 
provide an insight into the different processes of the exchange of gaseous 
matter and energy between the galaxy and the intergalactic medium. There is a 
permanent mass circulation between the Milky Way disk and the halo via the 
Galactic fountain process \citep{shapirofield76, bregman80, shapiro_benjamin91} 
where supernova explosions expel material out of the disk into the halo,
where it cools and then falls back onto the Galactic plane.
This gas enriches the circumgalactic environment with heavy elements. Also 
interactions with satellite galaxies inject interstellar material into the halo 
of the Milky Way \citep[e.g.,][]{mathewson74}. Finally, low-metallicity gas most 
likely is being accreted from the intergalactic medium (IGM; e.g., Wakker et 
al.\,1999; Richter et al.\,2001). These processes suggest that the gaseous 
material in the Milky Way halo is tightly connected to the on-going formation 
and evolution processes of our galaxy.

A large fraction of the neutral gas in the halo is in the form of clouds with 
radial velocities inconsistent with a simple model of Galactic rotation 
\citep[e.g.,][]{lockman02, lockman_murphyetal02}, the so-called intermediate- 
and high-velocity clouds  \citep[IVCs, HVCs,][] {mulleroortraimond63}. 
IVCs are most likely located in the lower Galactic halo at $d< 2$\,kpc, whereas 
HVCs are thought to be located in the outer halo of galaxies at $5<d<50$\,kpc 
\citep[e.g.,][]{Sembachetal91, vanWoerden99, wakker01, thom2006, 
wakker_york_howketal07, wakkeryorkwilhelmetal08}.

Many IVCs and HVCs most likely have different origins, as expected from the 
various gas circulation processes in the halo outlined above.  
The bulk of IVCs are located close to the disk and appear to fall towards the 
Galactic plane, which could be explained by the Galactic Fountain model 
\citep{shapirofield76, bregman80, shapiro_benjamin91}. 
Some of the HVCs most likely emerge during interaction processes between the 
Milky Way and satellite galaxies due to merging and accretion. One prominent 
example is the Magellanic Stream, which appears to be a tidal feature from the 
Magellanic Clouds as they interact with the Milky Way 
\citep[e.g.,][]{mathewson74}. HVCs also have been brought into relation with 
primordial building blocks of galaxies in a hierarchical formation scenario 
\citep[e.g.,][]{oort66, blitzspergeletal1999}.

A powerful method to analyse the characteristics of IVC and HVC gas at low 
redshifts is the analysis of high-resolution ultraviolet (UV) and optical 
absorption spectra together with \ion{H}{i} 21-cm emission data \citep{richterp06, 
benbekhtietal08}. The combination of these data provides important information 
about the spatial distribution and the physical properties of the clouds. 

This paper is based on our previous work \citep{benbekhtietal08}, in which we 
have analysed VLT/UVES (Ultraviolet and Visual Echelle Spectrograph) data of 
low-column density structures in the Milky Way halo. We have detected 
\ion{Ca}{ii} $\lambda\lambda 3934.8,3969.6$ (and in some cases also
\ion{Na}{i} $\lambda\lambda 5891.6,5897.6$) absorption features at low, intermediate 
and high radial velocities along 35 sight lines (out of a sample of 103)  
towards high-redshift quasars (QSOs). For several of these sight lines 
counterparts in 21-cm \ion{H}{i} emission were found using the 100-m radio 
telescope in Effelsberg.

In this work we concentrate on the follow-up study of \ion{H}{i} 21-cm small-
scale structures in directions that were previously observed with UVES and the 
Effelsberg telescope.
\citet{richterwestmeierbruens05} observed one sight line (PKS\,1448$-$232) with 
particularly prominent high-velocity \ion{Ca}{ii} and \ion{Na}{i} absorption 
lines with UVES, Effelsberg, and the VLA. The VLA \ion{H}{i} map resolves the 
HVC into several compact, cold clumps.
Here, we present new high-resolution \ion{H}{i} data of four other absorbing 
systems out of the sample of 35 sight lines. Using these observations it is 
possible to estimate important physical and chemical properties, e.g., lower 
distance and pressure limits, as well as \ion{Ca}{ii} and \ion{Na}{i} abundances.

Our paper is organised as follows. In Section\,\ref{secdata} we describe the 
data acquisition and reduction. A presentation of the four sight lines observed 
with VLA and WSRT is given in Section\,\ref{secsightlines}. In 
Section\,\ref{secdiscussion}, results regarding the properties of the observed 
small-scale structures are discussed. We conclude our study in 
Section\,\ref{secsummary}.

\section{Data acquisition and reduction}\label{secdata}

A detailed description of our UVES data and the initial \ion{H}{i} 21-cm 
observations with the 100-m telescope at Effelsberg is provided in 
\citet{benbekhtietal08}.

For four sight lines (QSO\,J0003$-$2323, QSO\,B1331$+$170, QSO\,B0450$-$1310 and 
J081331$+$254503) additional high-resolution data were obtained with the Very 
Large Array (VLA) and the Westerbork Synthesis Radio Telescope (WSRT). These 
data were used to search for small-scale structure within the diffuse gas clouds. 
Detailed information about the observations with both telescopes is provided in 
Table\,\ref{tab_observing_log}, showing the instrument used, the date of 
observation, the chosen configuration, the coordinates of the sources, field of 
view ($\Omega_\mathrm{HPBW}$), the spatial ($\Theta_\mathrm{HPBW}$) and instrumental spectral 
resolution ($\delta v$), the bandwidth ($\Delta \nu$), and the resulting noise 
 level ($\sigma_\mathrm{rms}$) in the final datacubes. 

\begin{table*}
\caption[Observation log]{Observational parameters of the high resolution measurements 
with the WSRT and the VLA. }
\label{tab_observing_log}
\small
\centering
\begin{tabular}{cccccccccc}\hline\hline
\rule{0pt}{3ex}Instrument&Date&Configuration&$l$&$b$&$\Omega_\mathrm{HPBW}$&$\Theta_\mathrm{HPBW}$&$\delta v$&$\Delta \nu$&$\sigma_\mathrm{rms}$\\
\rule{0pt}{3ex}&[m:y]&&[$\deg$]&[$\deg$]&[$\deg$]&[$\arcsec$]&[km\,s$^{-
1}$]&[MHz]&$[\mathrm{mJy\,Beam}^{-1}]$\\[1ex]\hline
\rule{0pt}{3ex}WSRT&Feb. 2007&$2 \times 48$&348.5&75.8&0.5&$80 \times 180$ 
&0.5&2.5&9.9\\[1ex]
&Feb. 2007&$2 \times 48$&196.9&28.6&0.5&$80 \times 120$&0.5&2.5&8.5 
\\[1ex]\hline
\rule{0pt}{3ex}VLA&Feb. 2007&DnC&49.4&-78.6&0.5&$100 \times 100$&1.3&1.56&10.6 
\\[1ex]
&Feb. 2007&DnC&211.8&-32.1&0.5&$100 \times 100$&1.3&1.56&13.3 \\[1ex]\hline
\end{tabular}
\end{table*}

\subsection{Westerbork Synthesis Radio Telscope (WSRT)}

The fourteen 25-m antennas of the WSRT were used in the $2\times48$ 
configuration, providing a good compromise between sensitivity and resolution. 
For our observations a correlator bandwidth of 2.5\,MHz was chosen. The spectral 
channel width is $0.5\,\mathrm{km\,s}^{-1}$ with a total of 1024 channels. The 
observing time for both clouds was 12\,h and additionally 1\,h for the (primary) 
pre- and post-calibrators.

The WSRT data were analysed using the MIRIAD software package. The first step 
was to flag bad data points in the source and calibrator datasets, in order to 
exclude these during the following reduction tasks. The primary calibrators were 
used for the absolute flux, phase, gain, and bandpass calibration. Due to  phase 
changes during the 12 h observation an additional relative phase calibration was 
necessary. For this purpose self-calibration was performed. Finally, dirty image 
data cubes were calculated via Fast Fourier Transformation (FFT) using a robust 
parameter of $r=0.5$. The cubes were gridded such that they have a velocity width of $0.5\,\mathrm{km\,s}^{-
1}$ per spectral channel. The CLEAN algorithm \citep{hoegbom74} was applied to 
reconstruct the image. The resulting data cubes and moment-0 maps were primary-
beam-corrected. 

\subsection{Very Large Array (VLA)}

The VLA observations were carried out using the twenty-nine 25-m radio antennas 
in the DnC configuration.
This hybrid configuration consists of the south-eastern and south-western arms 
in the smallest (D) configuration and the northern arm in the larger C 
configuration.
It is especially well suited for extended sources with a declination south of 
$\delta=-15\degr$. For each of the two polarisations the correlator provides a 
bandwidth of 1.56\,MHz with 256 spectral channels resulting in a channel 
separation of 1.3\,km\,s$^{-1}$.

The observing time for the sightline in the direction of QSO\,J0003$-$2323 was 
6\,h and 12\,h for QSO\,B1331$+$170 with additional 1\,h  allocated for the 
primary and secondary calibrators. The primary calibrators 0137$+$331 and 
0542$+$498 were observed for about 30\,min each at the beginning and the end of 
the observation periods. The secondary calibrators 0025$-$260 and 0447$-$220 
were observed once every hour for about 5 minutes.

The VLA data of QSO\,B0450$-$1310B and QSO\,J0003$-$2333 were contaminated by 
many interferences and antenna failures. As both sources were observed on 
multiple occasions, the individual parts had to be handled separately and were 
merged only after self-calibration. The data reduction was performed with MIRIAD. 
First, a calibration based on the observed primary and secondary calibrators was 
done. A median filter was applied to the visibilities per baseline to remove 
outliers, and three malfunctioning antennas as well as obvious interferences were 
completely flagged. Along with continuum subtraction, an iterative self-calibration was 
carried out to account for phase changes during the observations. After merging 
the respective visibility sets, data cubes were calculated via FFT using a 
robust parameter of $r=0.4$ for QSO\,J0003$-$2333 and of $r=2$ for QSO\,B0450$-
$1310B.
In order to improve sensitivity, the beam was artificially increased to a size 
of $100\arcsec\times 100\arcsec$ by convolution with a Gaussian. The resulting 
data cubes have a velocity width of $1.3\,\mathrm{km\,s}^{-1}$ per spectral 
channel. Again, primary beam corrections were applied.

\begin{figure*}
\centering
\includegraphics[width=0.85\textwidth,bb=20 86 985 840,clip=]{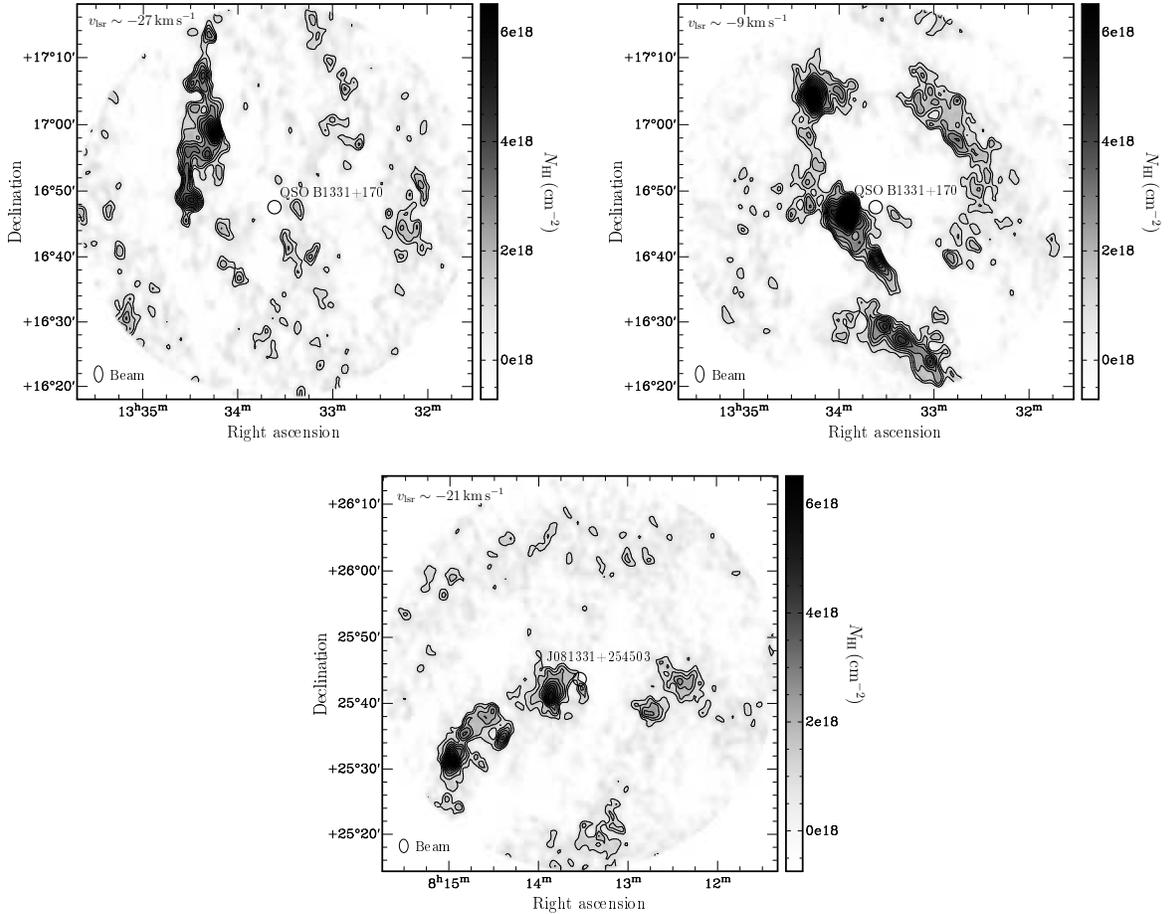}
\caption{\textbf{Upper panels}: WSRT 21-cm column density maps of the 
intermediate-velocity gas in the direction of the quasar QSO\,B1331+170. Several 
small \ion{H}{i} clumps with low column densities $(N_\mathrm{HI} \leq 
1.6\cdot10^{19}$\,cm$^{-2})$ are detected at two distinct velocities of 
$v_\mathrm{lsr}\approx-27$ and $v_\mathrm{lsr}\approx-10$\,km\,s$^{-1}$ (column densities 
are calculated using the velocity interval $v_\mathrm{lsr}=-30 \ldots -21$\,km\,s$^{-1}$ 
and $v_\mathrm{lsr}=-14 \ldots -1$\,km\,s$^{-1}$, respectively). The beam 
size of $80'' \times 150''$ is shown in the maps. The contours start at 
$2\sigma_\mathrm{rms}$ in steps of $\sigma_\mathrm{rms}=5 \cdot 10^{17}$\,cm$^{-
2}$. The  size of the synthesised beam is shown in the lower left corner.
\textbf{Lower panel}: WSRT 
21-cm map (integrated over $v_\mathrm{lsr}=-25 \ldots -18$\,km\,s$^{-1}$)  
of the intermediate-velocity gas in the direction of the quasar 
J081331$+$254503. Several small \ion{H}{i} clumps with low column densities 
$(N_\mathrm{HI} \leq2 \cdot 10^{19}$\,cm$^{-2})$ are detected. The beam size of 
$80'' \times 120''$ is shown in the figure. The contours start 
at $2\sigma_\mathrm{rms}$ in steps of $\sigma_\mathrm{rms}=5 \cdot 
10^{17}$\,cm$^{-2}$. Note, that the maps were not primary-beam corrected to improve visualisation.}
\label{fig_wsrt}
\end{figure*}

\begin{figure*}
\centering
\includegraphics[width=0.9\textwidth,bb=14 460 985 836,clip=]{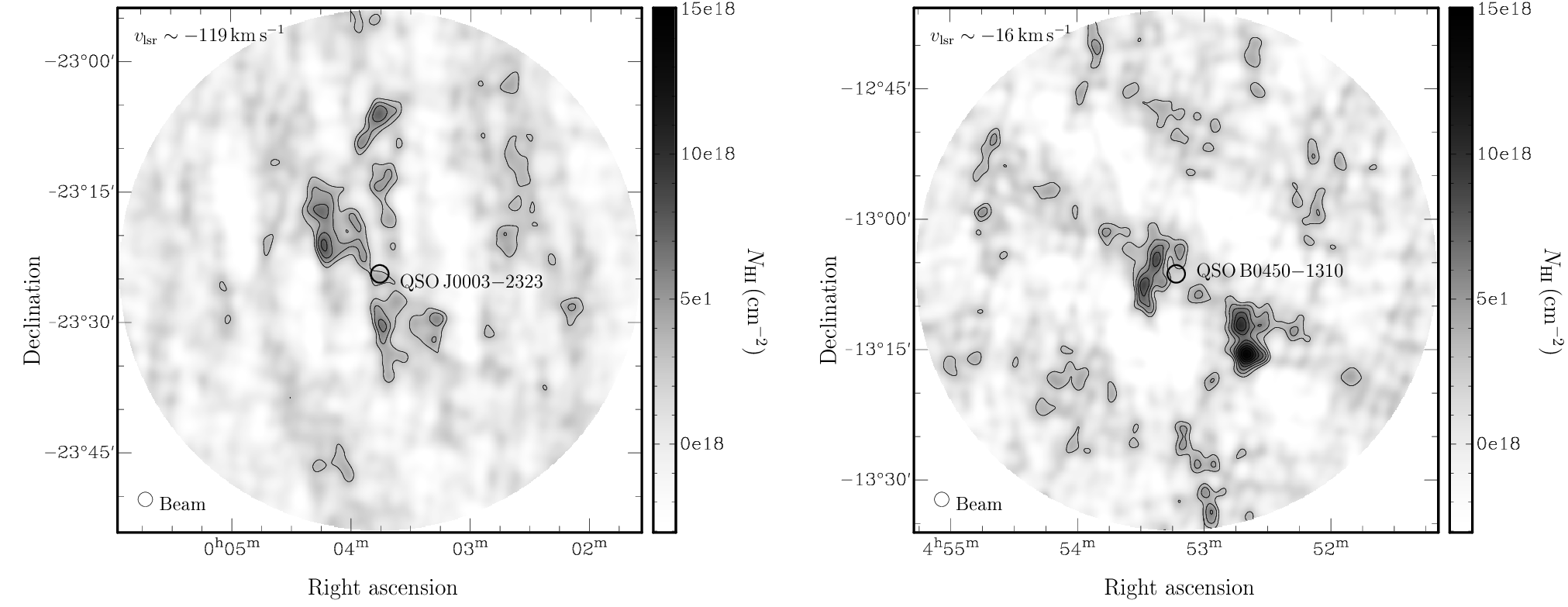}
\caption{\textbf{Left panel}: VLA 21-cm column density map (integrated over 
$v_\mathrm{lsr}= -121 \ldots -116$\,km\,s$^{-1}$) of the high-velocity gas in 
the direction of the quasar QSO\,J0003$-$2323. Several \ion{H}{i} clumps with 
low column densities $(N_\mathrm{HI} \leq 1.9 \cdot 10^{19}$\,cm$^{-2})$ are 
detected. The beam size of $100\arcsec \times 100\arcsec$ is indicated in the 
lower right corner. The contours start at $2\sigma_\mathrm{rms}$ in steps of 
$\sigma_\mathrm{rms}=1.5 \cdot 10^{18}$\,cm$^{-2}$.  \textbf{Right panel}: 
VLA 21-cm column density map (integrated over 
$v_\mathrm{lsr}=-22 \ldots -9$\,km\,s$^{-1}$) of the low-velocity 
gas observed along the sight line towards QSO\,B0450$-$1310. \ion{H}{i} clumps with column densities 
$N_\mathrm{HI} \leq 2.3 \cdot 10^{19}$\,cm$^{-2}$ are detected. The beam size of 
$100'' \times 100''$ is shown in the lower left corner. The contours start at 
$2\sigma_\mathrm{rms}$ in steps of $\sigma_\mathrm{rms}=1.5 \cdot 
10^{18}$\,cm$^{-2}$. Note, that the maps were not primary-beam corrected to improve visualisation.}
\label{fig_vla}
\end{figure*}

In Fig.\,\ref{fig_wsrt} we show the \ion{H}{i} (integrated) column densities of the QSO sightlines QSO\,B1331+170 and J081331+254503 associated to our absorbers. Figure\,\ref{fig_vla} contains the maps for QSO\,J0003$-$2323 and QSO\,B0450$-$1310. The velocity ranges over which was integrated are given in the figure captions. Note, that the spectral widths of the sources are small, hence, we decided to not show position--velocity plots (or individual spectral maps). We have clipped the maps for radii larger than 0.5 degrees (which is the HPBW of the primary beam of the WSRT and VLA). To improve visualisation the maps shown in Fig.\,\ref{fig_wsrt} and \ref{fig_vla} were not corrected for the primary beam. Hence, they do not show true flux values. The clumps which have been analysed are presented in a subsequent section with primary beam correction applied. In case of the VLA observations, residual image artifacts (stripes) are visible in the maps. We carefully inspected the data cubes in adjacent (empty) velocity planes. There, non of these stripes are present. It might be, that the \ion{H}{i} sources partially absorb the QSO continuum leading to baseline fitting problems. However, the clumps are detected with much higher significance than could be caused by the artificial emission. Nevertheless, the total column density of the clumps in direction of QSO\,J0003$-$2323 and QSO\,B0450$-$1310 may be overestimated by a few percent.

The datacubes were also inspected for emission at other radial velocities than that of the absorbers. In case of J081331+254503 we found few very weak ($N_\mathrm{HI} < 5 \cdot 10^{18}$\,cm$^{-2}$) small clumps at $v_\mathrm{lsr}\sim-60$\,km\,s$^{-1}$. It was not further analysed, because here we are interested in the correlation between emission and absorption only. The other sight lines do not reveal additional sources. 

In the next section we will present the results of the physical analysis of the clumps significantly detected.

\section{The selected sight lines}\label{secsightlines}
Inspired by the results of \citet{richterwestmeierbruens05} and 
\citet{benbekhtietal08} we re-observed four sightlines (out of a sample of 35) 
that exhibit prominent \ion{Ca}{ii} and/or \ion{Na}{i} halo absorption features 
with high-resolution radio interferometers. The aim of this study is to search 
for small-scale structures embedded in the neutral hydrogen gas that were 
possibly not resolved with the single-dish Effelsberg telescope. 

\begin{table*}[!tp]
\caption{Summary of UVES \ion{Ca}{ii}, \ion{Na}{i} and Effelsberg \ion{H}{i} 
measurements towards the four quasars.}
\label{tab_4_quasars}
\small
\centering
\begin{tabular}{ccccccccccc}\hline\hline
\rule{0pt}{3ex}Quasar&$l$&$b$&$v_\mathrm{LSR}$&$\log 
N_\mathrm{CaII}$&$b_\mathrm{CaII}$&$\log N_\mathrm{NaI}$&$b_\mathrm{NaI}$& $\log 
N_\mathrm{HI}$&$b_\mathrm{HI}$&HVC/IVC\\
\rule{0pt}{3ex}&[$\degr$]&[$\degr$]&[km\,s$^{-1}$]&[$N$ in cm$^{-2}$]&[km\,s$^{-
1}$]& [$N$ in cm$^{-2}$]&[km\,s$^{-1}$]&[$N$ in cm$^{-2}$]&[km\,s$^{-
1}$]&complex\\[1ex]\hline
\rule{0pt}{3ex} QSO J0003-2323&49.4&$-$78.6&$-$98& 11.9& 6 &-&-&-&-&MS\\
&&&$-$112& 11.8& 6 &-&-&19.6&19.3&MS\\
&&&$-$126& 11.9& 6 &-&-&-&-&MS\\\hline
\rule{0pt}{3ex} QSO B1331+170&348.5&75.8&$-$9&12.4&11&11.3&6&18.9&2.7&-\\
&&&$-$27&12.0&5&11.3&3&-&-&IV Spur\\\hline
\rule{0pt}{3ex} QSO B0450-1310B &211.8&$-$32.1& $-$5&-&-&11.5&5&19.8&2.3&- \\
&&&$-$20&-&-&12.0&2&19.3&6.8&- \\\hline
\rule{0pt}{3ex} J081331+254503&196.9&28.6&$-$23&-&-&11.1&5&19.8&37.0&(LLIV Arch)\\\hline
\end{tabular}
\end{table*}

In this section, the results of the high-resolution radio observations of the 
four \ion{Ca}{ii} and \ion{Na}{ii} halo absorbers are presented. We also summarise briefly the 
results of the analysis of the \ion{Ca}{ii} and \ion{Na}{i} (UVES) and low-
resolution \ion{H}{i} (Effelsberg) data. Voigt profiles were fitted to the 
absorption lines, Gaussians were used to parametrise the emission lines. 
Table\,\ref{tab_4_quasars} lists the position, LSR velocities, 
\ion{Ca}{ii}, \ion{Na}{i}, and \ion{H}{i} column densities, Doppler parameters $b$, 
and the possibly associated IVC or HVC complexes. All of the UVES and Effelsberg 
spectra shown here are plotted in the local standard of rest (LSR) velocity 
frame. Most of the gas in emission and absorption near zero
velocities belongs to the local Galactic disk. As described in 
\citet{benbekhtietal08}, the decision whether the gas is connected to the 
Galactic disk or halo is made based on a Milky Way model developed by 
\citet{kalberla03, kalberla07}. The model predicts for each direction the 
expected velocities for interstellar gas participating in Galactic disk rotation. 
Hence, the deviation velocity \citep{wakker91} can be calculated and is used to 
classify the sources as IVCs or HVCs.  The $4\sigma$ and $2\sigma$ detection limits for the \ion{Ca}{ii}~$\lambda 3934.77$ and \ion{Ca}{ii}~$\lambda 3969.59$ lines, respectively, correspond to a $\log (N_\mathrm{CaII}/\mathrm{cm}^{-2})$ detection limit of $\approx 11$. The parameters of all detected absorption and emission components for 35 sight lines are listed in \citet{benbekhtietal08}.

%, for which we have carried follow-up observations using \ion{H}{i} high-resolution synthesis telescopes. The columns give the name of the quasars, the coordinates, the velocities in the LSR frame, the logarithm of the \ion{Ca}{ii} column densities, the Doppler parameter ($b$-value) for the \ion{Ca}{ii} lines, the logarithm of the \ion{Na}{i} column densities, the $b$-value for the \ion{Na}{i} lines, the \ion{H}{i} column densities (from Effelsberg data), the $b$-value for the \ion{H}{i} lines, and the possibly associated HVC or IVC complexes. 

\begin{table*}[!tp]
\caption[Summary of the high-resolution \ion{H}{i} measurements]{Physical properties of the \ion{H}{i} clumps in the direction of the quasars 
J081331+254503, QSO B1331+170 (observed with the WSRT), QSO\,J0003$-$2323 
and QSO\,B0450$-$1310B (observed with the VLA). }
\label{tab_WSRT_clumps}
\small
\centering
\begin{tabular}{cccccccccccc}\hline\hline
\rule{0pt}{3ex}Quasar&Clump&$\alpha (J2000)$&$\delta 
(J2000)$&$v_\mathrm{LSR}$&$\Delta 
v_\mathrm{FWHM}$&$b_\mathrm{HI}$&Amplitude&$N_\mathrm{HI}$&$T_\mathrm{B}$&$T_\mathrm{kin}^\mathrm{max}$&$\phi$\\
\rule{0pt}{3ex}&&[h:m:s]&[d:m:s]&[km\,s$^{-1}$]&[km\,s$^{-1}$]&[km\,s$^{-
1}$]&[mJy/Beam]&[cm$^{-2}$]&[K]&[K]&[$\arcmin$]\\[1ex]\hline
\rule{0pt}{3ex}QSO\,J0003$-$2323
&A&00:03:46&$-$23:05:55&$-117.6$&$3.0 \pm 0.4$&1.8&$86 \pm 10$&$1.9 \cdot 
10^{19}$&5.2&200&3.1\\
&B&00:04:16&$-$23:17:13&$-120.0$&$3.1 \pm 0.2$&1.9&$37 \pm 3$&$8.9 \cdot 
10^{18}$&2.2&210&3.1 \\
&C&00:04:14&$-$23:21:13&$-119.7$&$3.2 \pm 0.9$&1.9&$34 \pm 8$&$9.1 \cdot 
10^{18}$&2.0&220&3.4 \\
&D&00:03:45&$-$23:30:37&$-119.1$&$5.6 \pm 1.5$&3.4&$17 \pm 4$&$7.3 \cdot 
10^{18}$&1.0&690&2.5 \\
&E&00:03:17&$-$23:29:43&$-118.6$&$2.3 \pm 0.3$&1.4&$33 \pm 3$&$6.9 \cdot 
10^{18}$&2.0&110&2.2\\\hline
\rule{0pt}{3ex}QSO\,B1331$+$170
&A&13:34:29&16:48:59&$-27.8$&$4.1 \pm 0.3$&2.5&$20 \pm 1$&$6.5 \cdot 
10^{18}$&1.0&370&2.9 \\
(component~1)&B&13:34:15&16:59:22&$-27.4$&$3.2 \pm 0.2$&1.9&$38 \pm 2$&$9.2 
\cdot 10^{18}$&1.9&220&2.0 \\
&C&13:34:30&17:06:42&$-27.5$&$5.7 \pm 2.4$&3.4&$18 \pm 4$&$8.2 \cdot 
10^{18}$&0.9&700&1.7\\
&D&13:34:22&17:07:46&$-25.8$&$4.9 \pm 1.0$&3.0&$25 \pm 4$&$9.9 \cdot 
10^{18}$&1.3&520&2.2 \\
&E&13:34:29&17:14:17&$-26.6$&$2.5 \pm 1.3$&1.5&$20 \pm 9$&$7.5 \cdot 
10^{18}$&1.0&140&1.7\\
&F&13:34:18&17:14:03&$-26.5$&$1.8 \pm 0.5$&1.1&$50 \pm 13$&$1.2 \cdot 
10^{19}$&2.6&70&1.8\\
\rule{0pt}{3ex}QSO\,B1331$+$170
&G&13:33:03&16:24:12&$-9.8$&$6.2 \pm 0.8$&3.7&$42 \pm 4$&$1.6 \cdot 
10^{19}$&2.1&840&2.3 \\
(component~2)&H&13:33:20&16:27:30&$-10.6$&$4.0 \pm 0.3$&2.4&$41 \pm 3$&$1.1 
\cdot 10^{19}$&2.1&350&2.2 \\
&I&13:33:54&16:46:55&$-10.3$&$3.7 \pm 0.3$&2.2&$34 \pm 2$&$8.9 \cdot 
10^{18}$&1.7&300&3.4\\
&J&13:34:17&17:04:54&$-11.5$&$4.1 \pm 0.6$&2.5&$48 \pm 6$&$1.5 \cdot 
10^{19}$&2.4&360&3.9 \\\hline
\rule{0pt}{3ex}QSO\,B0450$-$1310B
&A&04:52:40&$-$13:15:40&$-17.6$&$7.6\pm0.6$&4.6&$36\pm2$&$2.3 \cdot 
10^{19}$&2.2&1250&2.7 \\
&B&04:52:43&$-$13:12:16&$-16.8$&$6.9\pm1.4$&4.2&$20\pm3$&$1.4 \cdot 
10^{19}$&1.2&1050&1.7 \\
&C&04:52:18&$-$13:12:58&$-15.7$&$6.6\pm2.0$&4.0&$19 \pm 5$&$9.9 \cdot 
10^{18}$&1.2&960&1.8 \\
&D&04:53:29&$-$13:07:40&$-16.8$&$13.0\pm3.4$&7.9&$9\pm1$&$9.2 \cdot 
10^{18}$&0.6&3700&1.5 \\
&E&04:53:23&$-$13:04:40&$-18.3$&$7.8 \pm 0.2$&4.7&$9 \pm 1$&$8.6 \cdot 
10^{18}$&0.6&1300&1.4 \\\hline
\rule{0pt}{3ex}J081331$+$254503
&A&08:14:59&25:31:15&$-20.8$&$6.1 \pm 0.7$&3.6&$42 \pm 4$&$2.0 \cdot 
10^{19}$&2.5&800&2.7 \\
&B&08:14:51&25:35:35&$-22.1$&$4.8 \pm 0.7$&2.9&$17 \pm 2$&$6.8 \cdot 
10^{18}$&1.0&490&1.9 \\
&C&08:14:25&25:34:50&$-21.2$&$2.4 \pm 0.2$&1.5&$26 \pm 2$&$5.8 \cdot 
10^{18}$&1.6&130&1.6 \\
&D&08:13:52&25:41:19&$-21.4$&$4.0 \pm 0.2$&2.4&$17 \pm 1$&$5.8 \cdot 
10^{18}$&1.0&350&2.4 \\
&E&08:12:44&25:38:52&$-23.4$&$3.6 \pm 0.9$&2.2&$10 \pm 2$&$3.5 \cdot 
10^{18}$&0.6&280&2.4 \\
&F&08:12:25&25:43:38&$-23.2$&$8.3 \pm 2.1$&5.0&$7 \pm 1$&$4.3 \cdot 
10^{18}$&0.4&1500&3.0 \\\hline
\end{tabular}
\end{table*}

Using the reduced high-resolution data cubes and the resulting column density 
maps, various physical parameters were derived. All structures above a threshold 
of $4\sigma_\mathrm{rms}$ (in the integrated column density maps) were considered in our analysis. From the moment-0 
maps the peak column densities were inferred as well as the angular sizes of 
individual clumps (using the Gaussian radial fit method of the \textit{Karma} 
task \textit{kvis}). The latter, however, can only be considered rough estimates, 
as the clumps are neither necessarily spherical nor must have Gaussian column 
density profiles. The spectra containing the highest total flux were extracted, 
and Gaussian velocity profiles were fitted. This returned peak fluxes (which can 
be converted to peak brightness temperatures, $T_\mathrm{B}$), velocity widths 
(leading to an upper temperature limit), and LSR velocities of each clump. Note, 
that velocity profile widths are not only given in FWHM but are also converted 
to $b$-values to allow for better comparison with the absorption lines. The 
derived physical properties of the clumps are summarised in 
Table\,\ref{tab_WSRT_clumps}, containing the clump IDs, the coordinates, the velocities in the LSR frame, the FWHM derived from a Gauss fit and associated $b$-values, the peak amplitudes, the peak \ion{H}{i} column densities (with a typical  error of about $10^{18}$\,cm$^{-2}$), the peak brightness temperatures, the upper temperature limits, and the mean radial sizes, $\phi$ (FWHM).

%The columns give the name of the quasars, 

\subsection{QSO\,J0003$-$2323}\label{QSOJ0003}

\begin{figure}
\centering
\includegraphics[width=0.4\textwidth,clip]{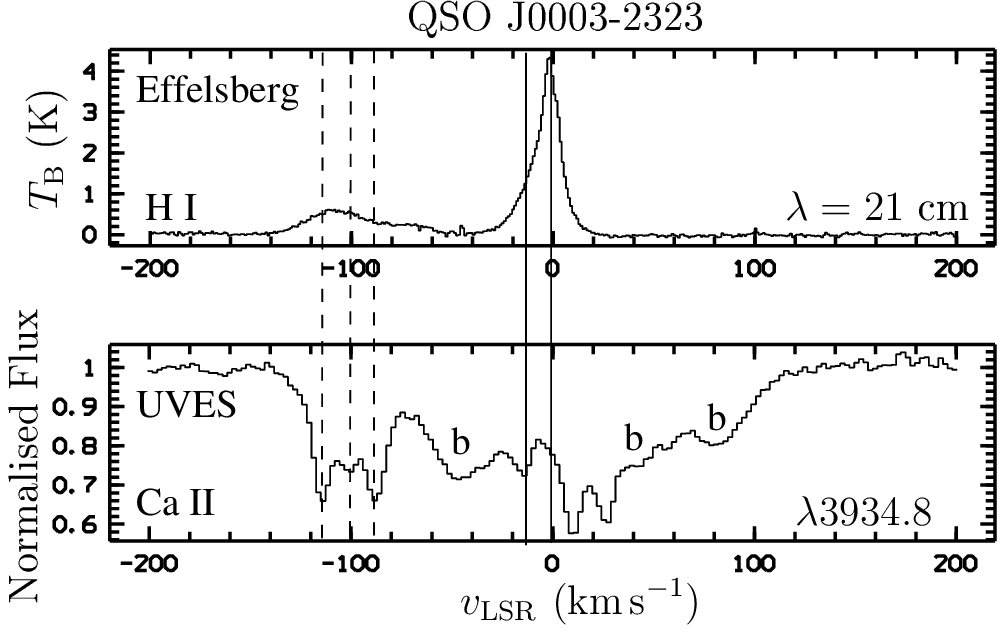}\\[1ex]
\includegraphics[width=0.45\textwidth,bb=11 295 517 
847,clip=]{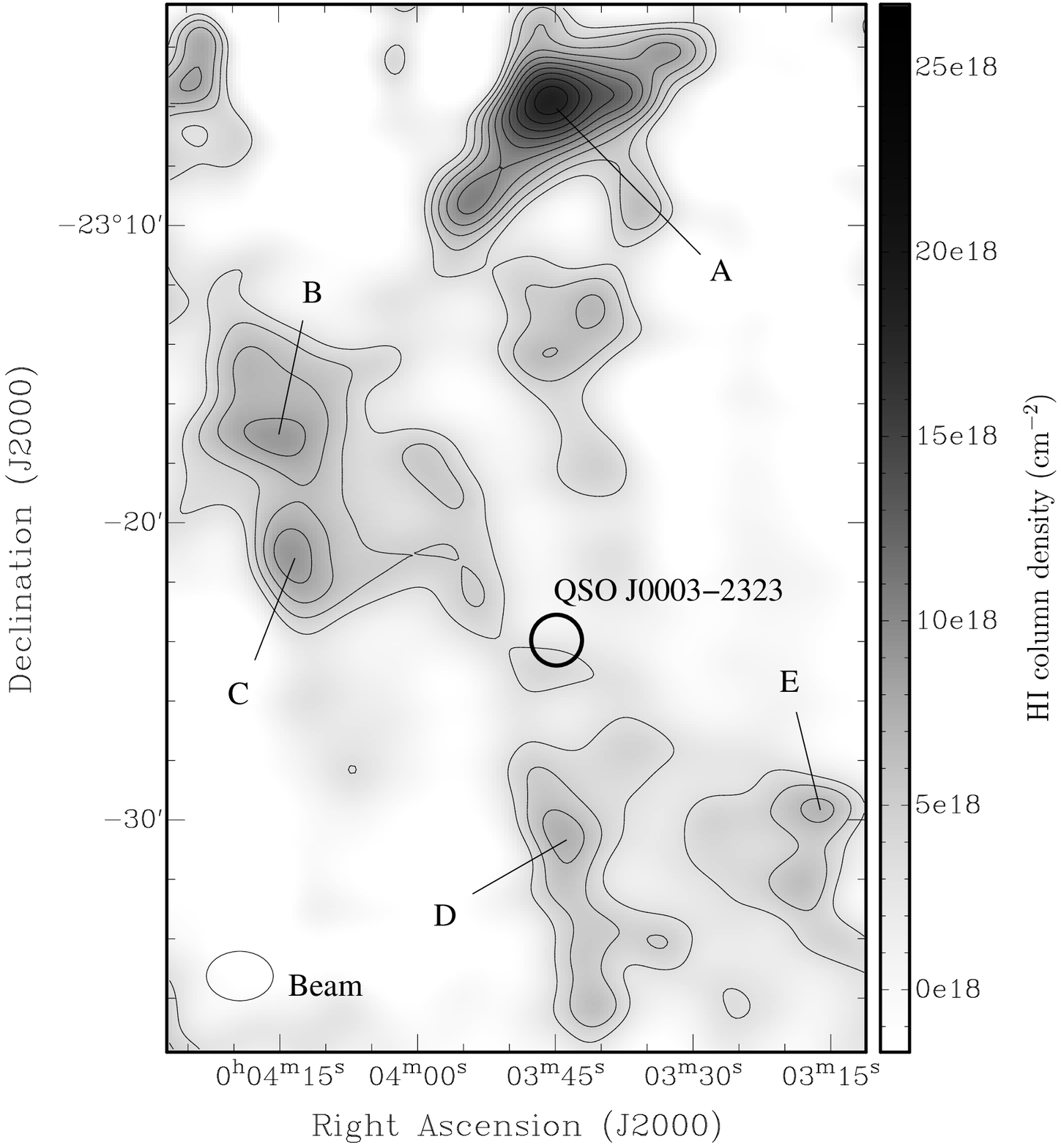}
\caption{\textbf{Upper panel}: \ion{Ca}{ii} $\lambda 3934.8$ absorption profile 
towards QSO\,J0003$-$2323. Additionally, the corresponding emission profile of 
\ion{H}{i} 21-cm is shown as observed with the 100-m telescope at Effelsberg. 
The dashed vertical lines indicate the absorption components which were identified. The 
solid lines mark the minimal and maximal LSR velocities which are expected for 
the Galactic disc gas in this direction according to the Milky Way model 
developed by \citet{kalberla03, kalberla07}; b denotes blending lines from 
intervening absorption systems.
\textbf{Lower panel}: VLA 21-cm column density map showing the clumps A to E in 
the direction of the quasar QSO\,J0003$-$2323. The contour lines are identical to those in Fig.\,\ref{fig_vla} (left panel).}
\label{fig_he0001}
\end{figure}

The upper panel of Figure\,\ref{fig_he0001} shows the multi-component 
\ion{Ca}{ii} halo absorbers in the direction of QSO\,J0003$-$2323 as observed 
with UVES. 
\ion{Ca}{ii} $\lambda \lambda ~3934.77, 3969.59$ absorption at high velocities 
is detected with high significance ($31 \sigma_\mathrm{rms}$) level, but no 
corresponding \ion{Na}{i} is seen. 
The \ion{Ca}{ii} absorption spreads between $v_\mathrm{lsr} \approx -120\ldots-
80$\,km\,s$^{-1}$ having $b_\mathrm{CaII}\approx 6$\,km\,s$^{-1}$. A 
corresponding broad ($b=19.3$\,km\,s$^{-1}$) 21-cm emission line is found in the 
Effelsberg data at $v_\mathrm{lsr} \approx -112$\,km\,s$^{-1}$ with an \ion{H}{i} 
column density of about $\log (N_\mathrm{HI}/\mathrm{cm}^{-2})=19.6$ (see upper 
panel of Fig.\,1).  This absorbing system is kinematically compact, i.e., 
multiple components with comparable $b$-values are spread over less than 
$\approx 100$\,km\,s$^{-1}$ \citep{dingcharltonchurchill05}. The position as 
well as the velocities of this \ion{Ca}{ii} absorber indicate a possible 
association with the Magellanic Stream \citep{benbekhtietal08}.

The lower panel of Figure\,\ref{fig_he0001} shows the high-resolution \ion{H}{i} 
column density map of the high-velocity gas in the direction of QSO\,J0003$-
$2323 integrated over $v_\mathrm{lsr}= -121 \ldots -116$\,km\,s$^{-1}$, 
as observed with the VLA.
The observations reveal the presence of several \ion{H}{i} clumps. Clumps with  
$\geq 4\sigma_\mathrm{rms}$ ($\sigma_\mathrm{rms}=1.6 \cdot 10^{18}$\,cm$^{-2}$) 
significance level are labelled with letters A to E. 

The small-scale clumps have \ion{H}{i} peak column densities of $\approx 7 
\ldots 9 \cdot 10^{18}$\,cm$^{-2}$, except for clump A, which has $1.9 \cdot 
10^{19}$\,cm$^{-2}$, corresponding to peak brightness temperatures of 
$T_\mathrm{B}\approx 1 \ldots 2$\,K (clump A: 5\,K).  All clumps have angular 
diameters of $\phi < 3.5\arcmin$ (FWHM) and are mostly relatively isolated or 
only weakly connected to each other. However, this could be an instrumental 
effect as the interferometric data lack small spatial frequencies. Note also, 
that the clumps might not be resolved by the VLA beam, as the spatial features 
visible in the map match approximately the size of the beam ($100\arcsec \times 
100\arcsec$). This means that there could be even more substructure within the 
cores.

The line widths are very small with $\Delta v_\mathrm{FWHM} \leq3 $\,km\,s$^{-
1}$, typically, implying that the absorbing gas is cool with an upper 
temperature limit of $T_\mathrm{kin} \approx 200$\,K. Only clump D reveals 
slightly larger velocity widths, leading to an upper limit of $T_\mathrm{kin} 
\leq 700$\,K.  An important point is that in the \ion{H}{i} single dish and 
interferometric data only one emission line was detected, while the absorption 
measurements reveal a multi-component structure. There also exists a slight 
velocity offset between the \ion{Ca}{ii} absorption components and the 
velocities of the \ion{H}{i} clumps seen in emission (see Tables 2 \& 3). 
This indicates that absorption and emission measurements trace somewhat 
different regions of the same overall structure.
Fig.\,1 (lower panel) shows that the quasar sightline do not pass any of the 
clumps directly, but rather traces the outer boundary of clump C, which is 
barely detected in 21-cm emission. The detection of multiple, relatively strong  
\ion{Ca}{ii} absorption components towards QSO\,J0003$-$2323 thus indicates that  
there are additional, low-column density neutral gas structures along this
sightlines that cannot be detected and/or resolved even with deep 21-cm synthesis 
observations. The complex spatial structure and kinematics of the outer layers 
of HVCs has been previously observed \citep{bruens98, bruenskerppagels01, 
benbekhti06} and was also found in numerical simulations \citep{quilismoore01}, 
pointing towards interactions of the clouds with the ambient medium, i.e., ram-
pressure causing head-tail structures. 

\begin{figure*}
\centering
\begin{minipage}[b]{0.45\textwidth}
 \includegraphics[width=0.95\textwidth,clip]{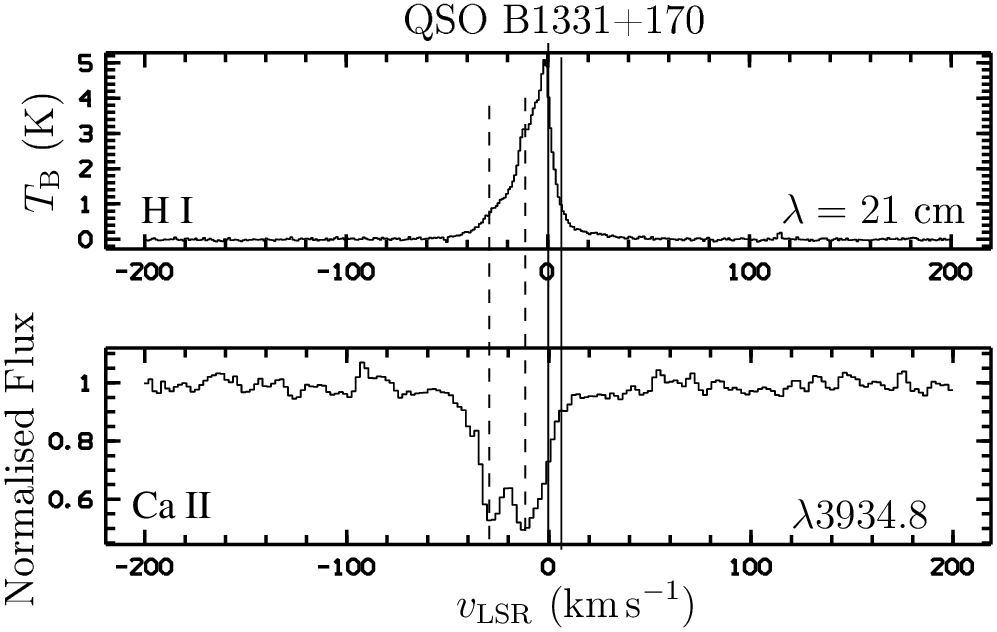}\\[1ex]
\includegraphics[width=\textwidth, bb=10 291 562 843, 
clip]{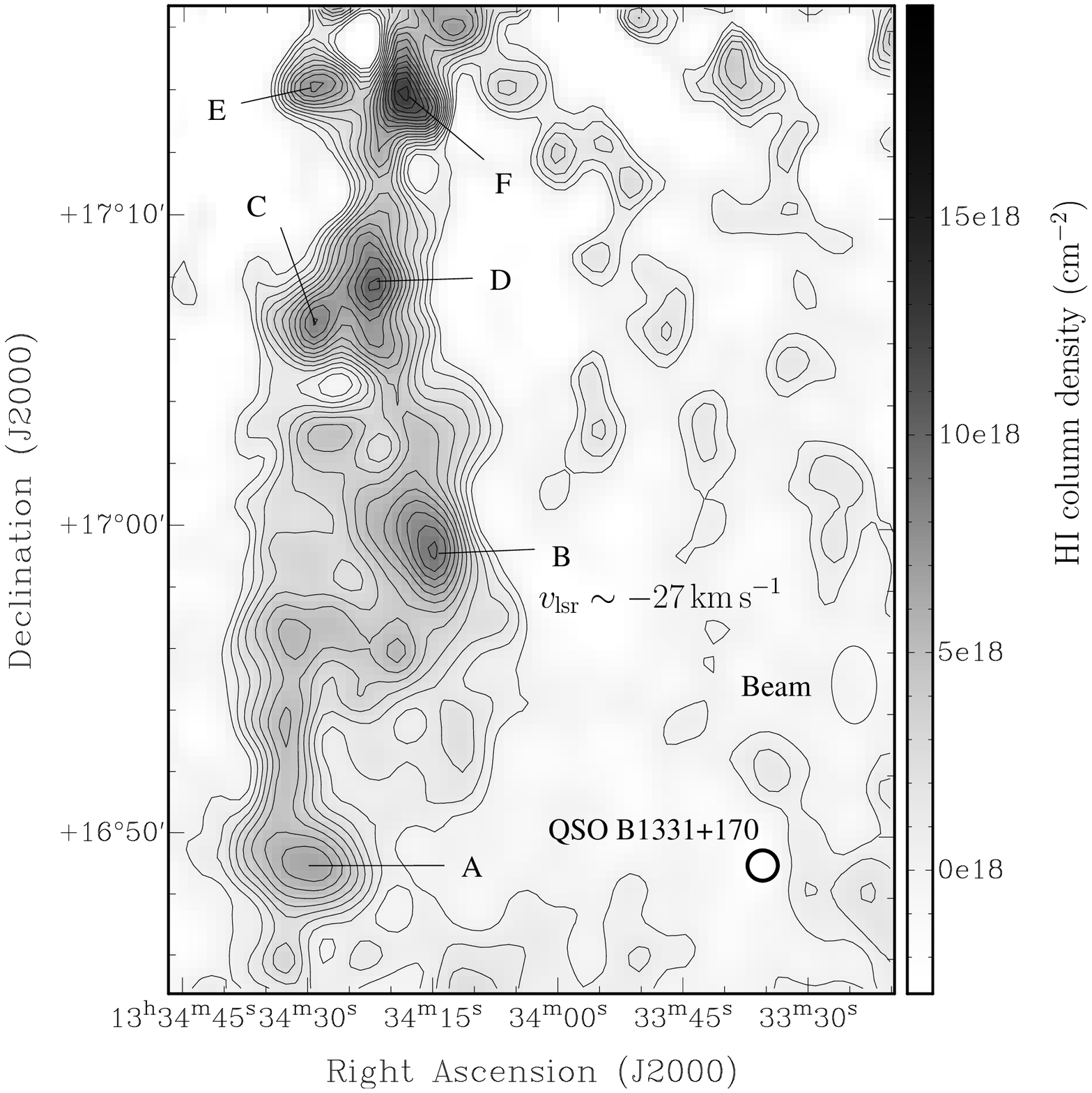}
\end{minipage} \hfill
\includegraphics[width=0.52\textwidth, bb=55 299 530 826, 
clip]{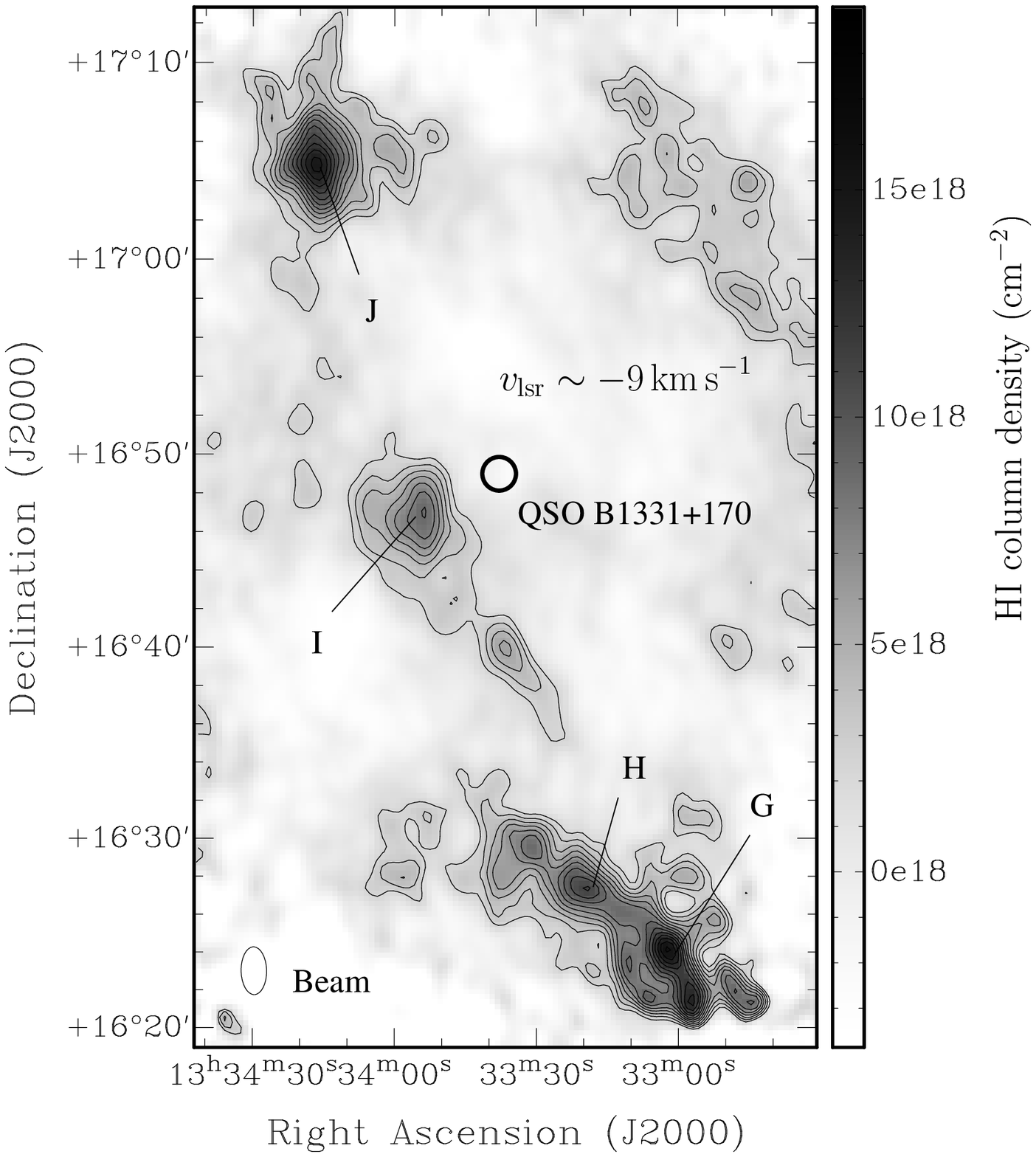}
\caption{\textbf{Upper left panel}: \ion{Ca}{ii} absorption and corresponding 
\ion{H}{i} emission spectra towards QSO\,B1331$+$170.
\textbf{Lower left  and right panel}: WSRT 21-cm column density maps of the 
intermediate-velocity gas in the direction of the quasar QSO\,B1331+170. 
Note, that for the component at $v_\mathrm{lsr}\approx-27$ km\,s$^{-1}$ the 
line of sight towards QSO\,B1331+170 
 lies about $10\arcmin$ away from the 
nearest clump, while for the second component the line of sight is located near 
clump~I. The contour lines are identical to those in Fig.\,\ref{fig_wsrt} (upper panels). } 
\label{fig_QSOB1331_2komps}
\end{figure*}

\subsection{QSO\,B1331$+$170}

The intermediate-velocity \ion{Ca}{ii} absorber towards 
QSO\,B1331$+$170 is shown in the upper left panel of 
Figure\,\ref{fig_QSOB1331_2komps}.
This two-component system is detected in \ion{Ca}{ii} (at $19 \sigma$ 
significance), as well as in \ion{Na}{i}. The lines spread between 
$v_\mathrm{lsr} \approx -40\ldots0$\,km\,s$^{-1}$. The position as well as the 
velocity of the line at $v_\mathrm{lsr} \approx -27$\,km\,s$^{-1}$ indicate a 
possible association with the intermediate-velocity complex IV Spur.

There is a corresponding \ion{H}{i} line at $v_\mathrm{lsr} \approx -
9$\,km\,s$^{-1}$ detected with the 100-m telescope at Effelsberg and possibly 
there is also a feature at $-27$\,km\,s$^{-1}$, however, not detected clearly. 
The peak column density of the observed line is $N_\mathrm{HI}=6.3 \cdot 
10^{19}$\,cm$^{-2}$. 

The high-resolution WSRT 21-cm column density map (integrated over $v_\mathrm{lsr}=
-30 \ldots -21$\,km\,s$^{-1}$ and $v_\mathrm{lsr}=-14 \ldots -1$\,km\,s$^{-1}$, respectively)
 reveals the presence of two 
components at $v_\mathrm{lsr}\approx-27$ km\,s$^{-1}$ (component~1) and 
$v_\mathrm{lsr}\approx-10$ km\,s$^{-1}$ (component~2) nicely matching the 
absorption signatures in velocity. In the first case, six clumps are identified, 
labelled with letters A to F. The \ion{H}{i} peak column densities range from 
about $\approx 6.5 \cdot 10^{18}$\,cm$^{-2}$  to $1.2 \cdot10^{19}$\,cm$^{-2}$ 
and have peak brightness temperatures on the order of $T_\mathrm{B}\approx 1.0 
\ldots 2.5$\,K. Compared to the other three sight lines and component~2, this 
structure is spatially more compact. Furthermore, the individual clumps are less 
isolated. For component~1 the spatial separation between the position of the 
quasar and the high-resolution emission is about $10\arcmin$. The angular 
diameter $\phi$ of the clumps is about $2\arcmin$. The line widths of the six 
clumps are in the range of  $\Delta v_\mathrm{FWHM}\approx2 \ldots 6$\,km\,s$^{-
1}$ which implies that the absorbing gas has a temperature of 
$T_\mathrm{kin}^\mathrm{max}\leq 700$\,K. Clump~F has an extraordinarily small 
line width of only 1.8\,km\,s$^{-1}$ ($T_\mathrm{kin}^\mathrm{max}\leq 70$\,K). 
Component~2 exhibits four clumps (G to J) having larger column densities of 
$\approx 0.9\ldots1.6 \cdot 10^{19}$\,cm$^{-2}$ and peak brightness temperatures 
($\approx2$\,K). 

\begin{figure}
\centering

 \includegraphics[width=0.45\textwidth,bb=0 0 540 539,clip=]{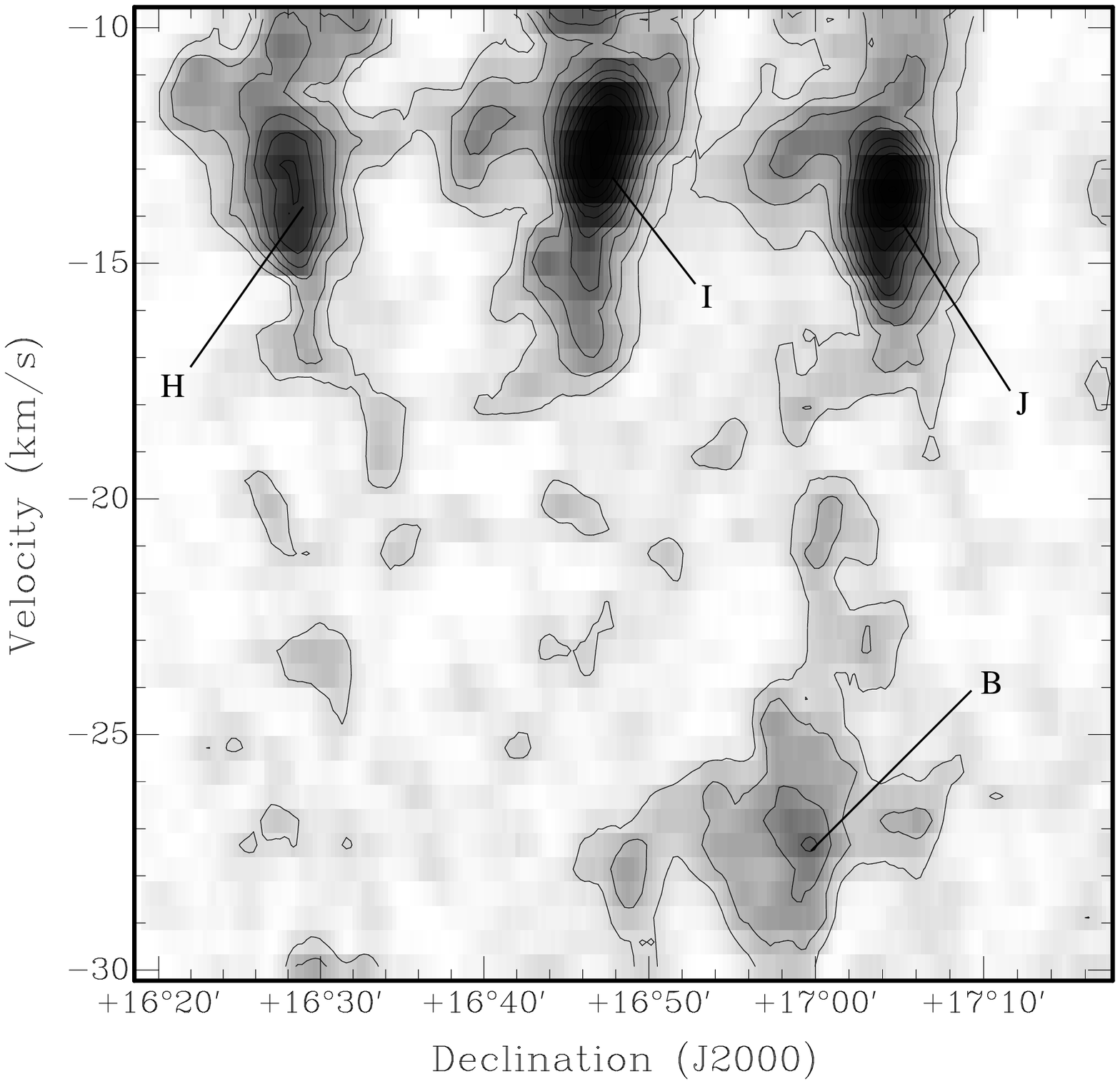}
\caption{Position--velocity map of QSO\,B1331+170 integrated over the range $\mathrm{R.A.}=[13^\mathrm{h}32^\mathrm{m}31^\mathrm{s},13^\mathrm{h}35^\mathrm{m}38^\mathrm{s}]$. We did not use the primary-beam corrected cube as the high noise levels towards the edges would degrade the signal-to-noise level in the map significantly. Milky Way emission was not included in the image section. Contours start at $2\sigma_\mathrm{rms}$ in steps of $\sigma_\mathrm{rms}$. The map reveals a faint emission feature between clump B (which is part of component 1) and clump J (part of component 2). However, it is not sure, if there is a physical connection or whether this signal is just noise.  } 
\label{fig_QSOB1331_bridge}
\end{figure}

To check whether there is a physical connection between both components we computed an integrated position--velocity map; see Fig.\,\ref{fig_QSOB1331_bridge}. There is a faint emission feature connecting clump~B and clump~J, however the feature is close to the noise level ($4\sigma_\mathrm{rms}$). Possibly, both components are due to an outflow, with component~2 being very close (in velocity) to the Milky Way emission. Component~1 might just represent a feature which has been expelled more recently (hence having a larger velocity).

\subsection{QSO\,B0450$-$1310}
\begin{figure}
\centering
\includegraphics[width=0.4\textwidth,clip]{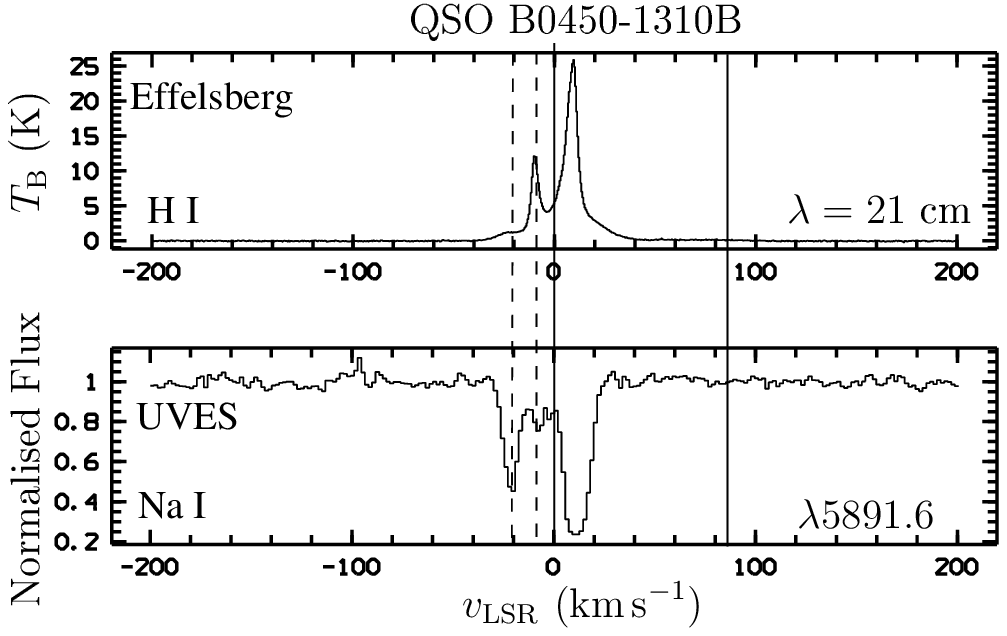}\\[1ex]
\includegraphics[width=0.45\textwidth,clip, bb=11 298 574 
844]{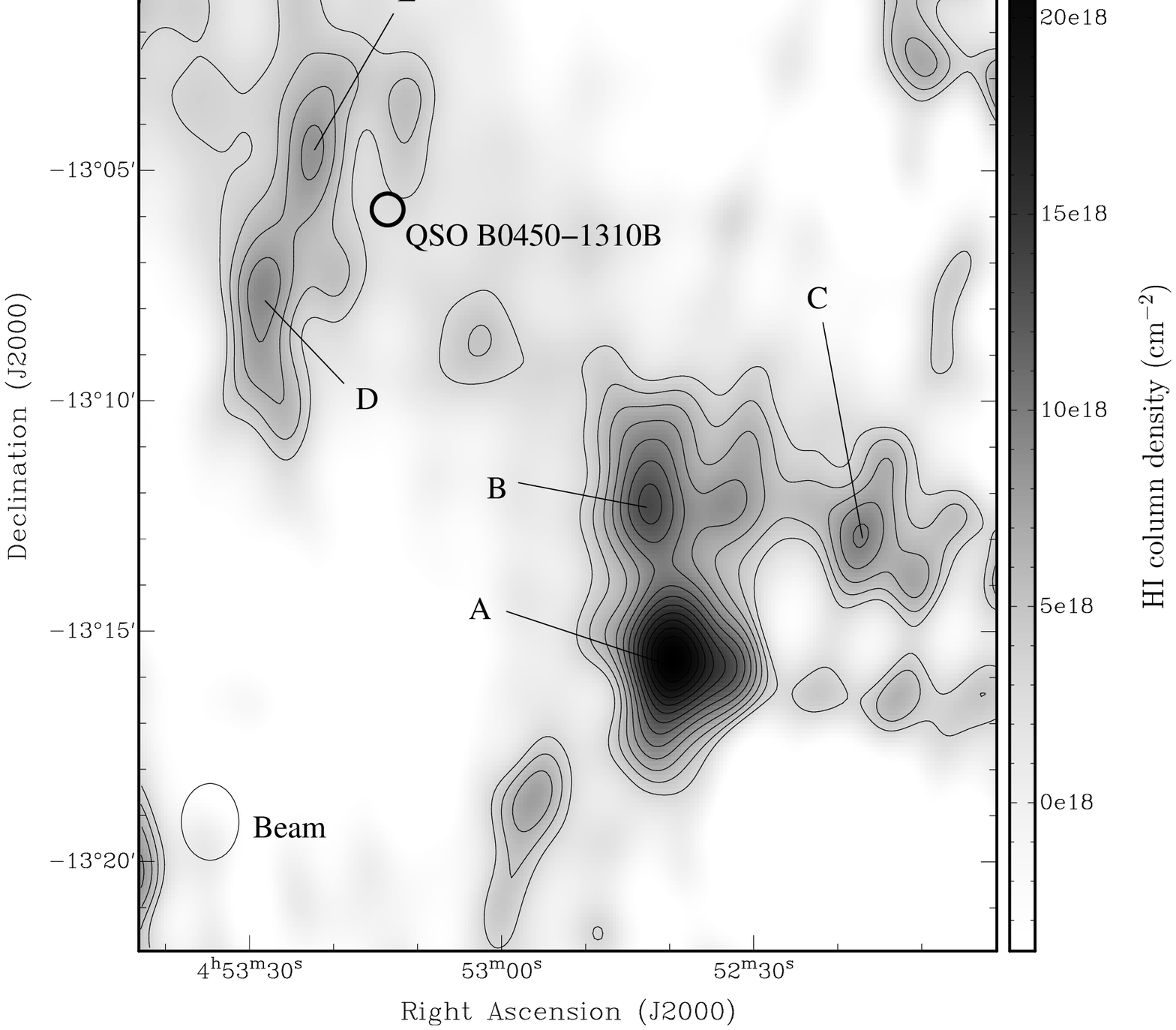}
\caption{\textbf{Upper panel}: \ion{Na}{i} absorption spectra towards 
QSO\,B0450$-$1310B obtained with UVES. There is no \ion{Ca}{ii} spectrum 
available for this sight line. The corresponding Effelsberg \ion{H}{i} emission 
line profile is shown.
\textbf{Lower panel}: VLA 21-cm column density map of the low-velocity 
gas observed towards QSO\,B0450$-$1310B. The sightline passes the outer parts of clumps D and E. 
The contour lines are identical to those in Fig.\,\ref{fig_vla} (right panel).}
\label{fig_QSOB0450}
\end{figure}

Figure\,\ref{fig_QSOB0450} shows the two-component absorber at low and 
intermediate velocities towards QSO\,B0450$-$1310. Two \ion{Na}{i} lines 
($\lambda \lambda ~ 5891.58, 5897.56$) were clearly detected at a $25 \sigma$ 
and $11 \sigma$ level, respectively. Unfortunately, no \ion{Ca}{ii} data 
currently are available for this line of sight. The two absorption lines have 
central velocities of $v_\mathrm{lsr} \approx -5$ and $-20$\,km\,s$^{-1}$. There 
is no obvious association with any known IVC complex.

For both components \ion{H}{i} counterparts in the spectrum obtained with the 
100-m telescope at Effelsberg were found, having small $b_\mathrm{HI}$-values of 
$2.3$ and $6.8$\,km\,s$^{-1}$.
Such narrow line widths were also measured for the \ion{Na}{i} absorption lines 
($b_\mathrm{NaI}=2.5$\,km\,s$^{-1}$). The column densities of both \ion{H}{i} 
systems are relatively small with $\log (N_\mathrm{HI}/\mathrm{cm}^{-2})=19.8$ 
and $\log (N_\mathrm{HI}/\mathrm{cm}^{-2})=19.3$. The \ion{Na}{i} absorption 
line at $-20\,\mathrm{km\,s}^{-1}$ reveals a larger column density and smaller 
width than the line at $-5\,\mathrm{km\,s}^{-1}$. Interestingly, this is the 
opposite of what is observed for the \ion{H}{i} emission.

The \ion{H}{i} high-resolution column density map (integrated over 
$v_\mathrm{lsr}=-22 \ldots -9$\,km\,s$^{-1}$) obtained with the VLA displays two main 
structures consisting of few clumps each. Clumps with a significance level of 
$\geq 4\sigma_\mathrm{rms}$ ($\sigma_\mathrm{rms}=1.6 \cdot 10^{18}$\,cm$^{-2}$) 
were labelled with letters A to E. The peak brightness temperatures are in the 
range of $T_\mathrm{B}= 0.6 \ldots 2.2$\,K, corresponding to \ion{H}{i} column 
densities between $9 \cdot 10^{18}$\,cm$^{-2}$  and $2.3 \cdot 10^{19}$\,cm$^{-
2}$. The angular diameters are very small with $\phi \approx 
1.5\ldots2.5\arcmin$. The line widths of the clumps vary from about $\Delta 
v_\mathrm{FWHM}=7$\,km\,s$^{-1}$ up to $\Delta v_\mathrm{FWHM}=13$\,km\,s$^{-1}$ 
which correspond to upper temperature limits of $T_\mathrm{kin} \approx 1000$\,K 
and $T_\mathrm{kin} \approx 3700$\,K, which is much higher than derived for the 
other sight lines. The small-scale features lie at velocities of $v_\mathrm{lsr} 
\approx -18\,\mathrm{km\,s}^{-1}$, which is close to the higher-column density 
absorption line, while the diffuse gas observed with the 100-m telescope is 
mainly found at different velocities. This shows again, that absorption and 
emission measurements sample somewhat different regions and gas phases within 
the overall cloud complex. Possibly, the gas interacts with an ambient medium, 
producing colder cores containing \ion{Na}{i} in the interface region, while the 
diffuse gas is dragged away by ram pressure.

\subsection{J081331$+$254503}
\begin{figure}
\centering
\includegraphics[width=0.4\textwidth,clip]{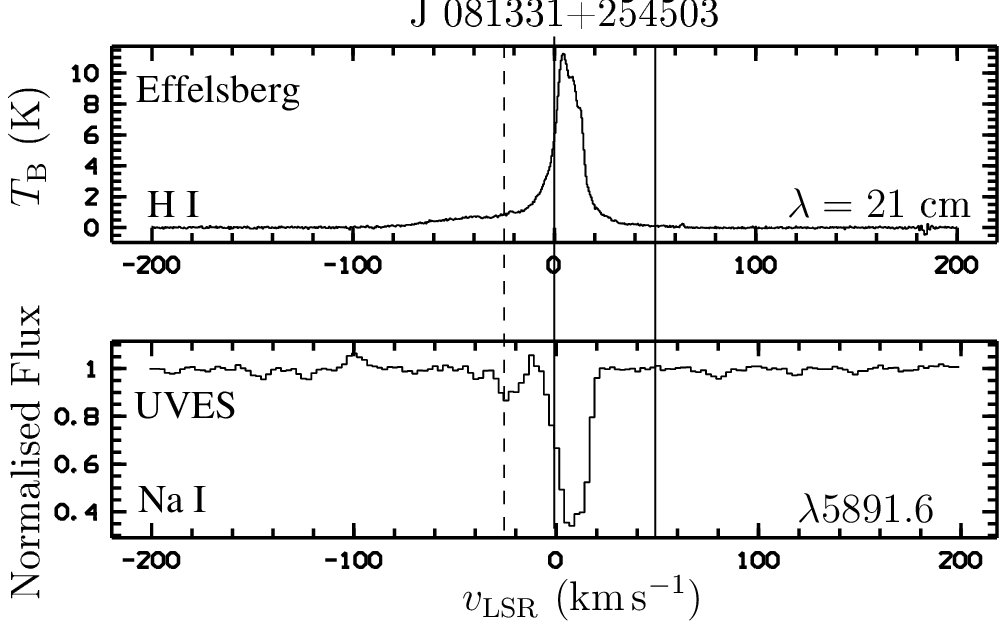}\\[1ex]
\includegraphics[width=0.5\textwidth, bb=11 529 568 
844]{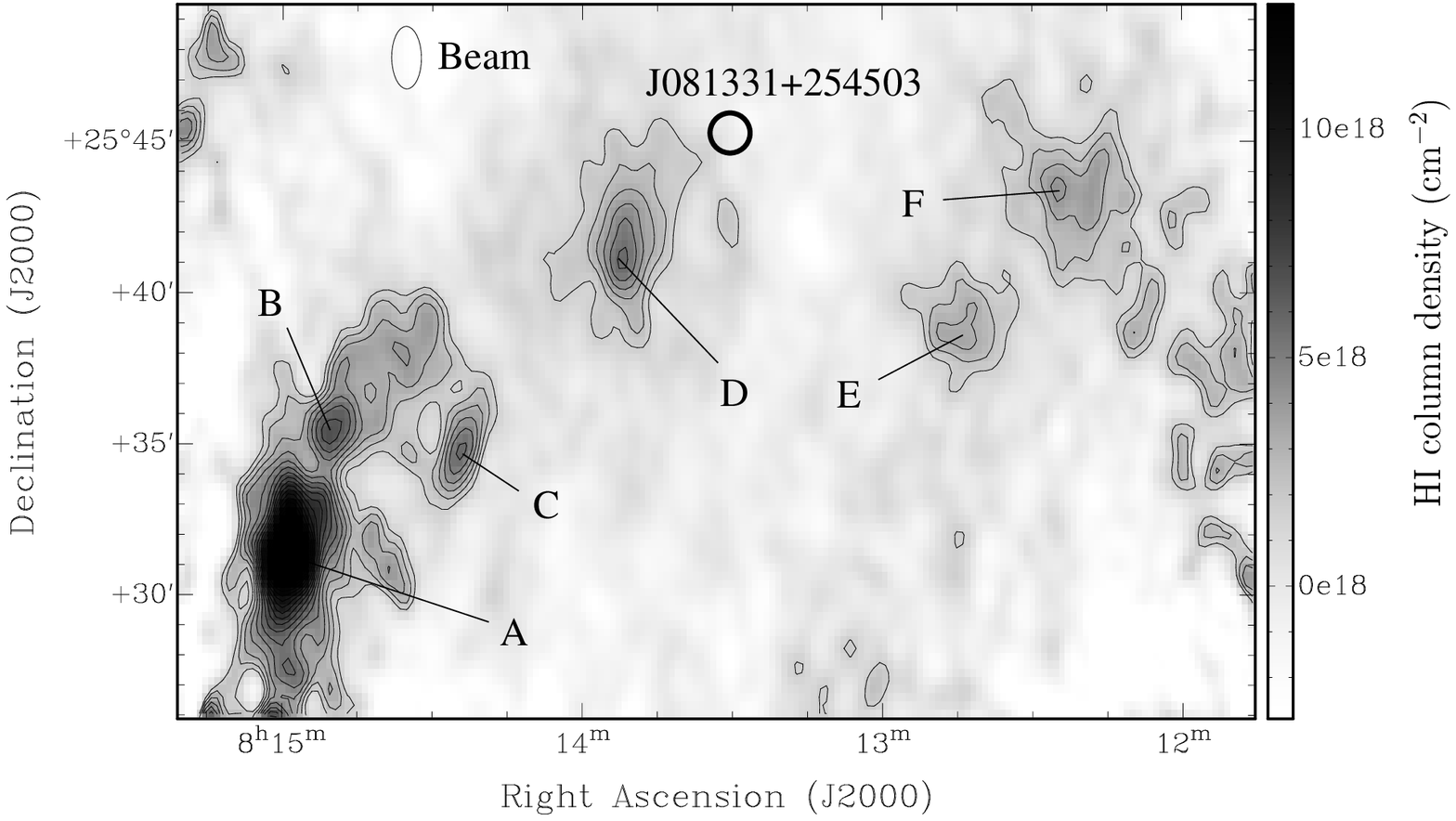}
\caption{\textbf{Upper panel}: \ion{Na}{i} absorption and corresponding 
Effelsberg \ion{H}{i} emission spectra in the direction of J081331$+$254503. No 
\ion{Ca}{ii} data are available for this sight line. \textbf{Lower panel}: WSRT 
21-cm map   
of the intermediate-velocity gas in the direction of the quasar 
J081331$+$254503. The line of sight towards J081331+254503 
 passes the very outer envelope of clump D.
The contour lines are identical to those in Fig.\,\ref{fig_wsrt} (lower panel).}
\label{fig_J081331}
\end{figure}
The only single-component absorber in our absorber sample lies in the direction 
of the quasar J081331$+$254503 (see Figure\,\ref{fig_J081331}). \ion{Na}{i} 
$\lambda\lambda 5891.58, 5897.56$ absorption is detected at $v_\mathrm{lsr} 
\approx -21$\,km\,s$^{-1}$ at a $11 \sigma$ level. There is no \ion{Ca}{ii} 
spectrum available for this sight line. The corresponding Effelsberg \ion{H}{i} 
emission line has a $b_\mathrm{HI}$-value of about $37.0$\,km\,s$^{-1}$ and a column density 
of  $N_\mathrm{HI}=6.3 \times 10^{19}$\,cm$^{-2}$.  There is no association with 
any known IVC complex.

Figure~\ref{fig_J081331} shows the WSRT 21-cm map of the intermediate-velocity 
gas in the direction of the quasar integrated over $v_\mathrm{lsr}=-25 \ldots -18$\,km\,s$^{-1}$. 
The high-resolution observations reveal the 
presence of six \ion{H}{i} clumps labelled with the letters A to F. These small-scale 
clumps ($\phi \leq 3\arcmin$) have typical peak \ion{H}{i} column 
densities in the range of $\approx 0.4 \ldots 2.0 \cdot 10^{19}$\,cm$^{-2}$ and 
peak brightness temperatures of $T_\mathrm{B}\approx 0.5\ldots2.5$\,K.  The line 
widths of  $\Delta v_\mathrm{FWHM} \leq 8.3$\,km\,s$^{-1}$ lead to an upper 
temperature limit of $T_\mathrm{kin} \approx 1500$\,K. The mean radial velocity 
is $v_\mathrm{lsr} \approx -22$\,km\,s$^{-1}$ matching that of the absorption.

It might be possible that the clumps are part of the low-latitude IV arch (LLIV arch). However, they have slightly smaller radial velocities than reported for the LLIV arch \citep{wakker01} so that the association is unsure. 

\section{Discussion}\label{secdiscussion}

As suggested by previous results \citep{richterwestmeierbruens05}, the follow-up 
observations with the WSRT and the VLA of four quasar sight lines that exhibit 
low-, intermediate- and high-velocity \ion{Ca}{ii} (\ion{Na}{i}) absorption in optical spectra reveal 
the presence of several compact \ion{H}{i} clumps that are associated with the 
absorbers. Note, that in the case of the low-velocity sources (QSO\,B1331+170 component\,2 and eventually QSO\,B0450$-$1310, both having $v_\mathrm{dev}<10\,\mathrm{km\,s}^{-1}$) it is unclear whether the sources are belonging to the Milky Way disc (at somewhat anomalous velocities due to turbulent motions) or tracing the interface region between the halo and the disc. All our clouds have \ion{H}{i} column densities below $3\cdot10^{19}\,\mathrm{cm}^{-2}$ which is lower than the median value reported by \citet{heiles_troland03} of $5\cdot10^{19}\,\mathrm{cm}^{-2}$ (CNM) for disk clouds in the solar neighbourhood and also lower than the column densities of clouds at high forbidden velocities in the inner galaxy as found by \citet{stiletal06}. On the other hand, \citet{fordetal08} find \ion{H}{i} clouds in the lower halo dominated by galactic rotation with  $N_\mathrm{HI}\sim0.2\ldots2.2\cdot10^{19}\,\mathrm{cm}^{-2}$ which is similar to our results. Hence, without knowing the distance to the sources we can not rule out one of the two possibilities.

In the following, we discuss the physical properties of these clumps.

\subsection{Distance-dependent physical parameters}

The coordinates as well as the radial velocities of QSO\,B1331$+$170 
(component~1) and QSO\,J0003$-$2323 suggest that the low-column density gas is 
part of the IVC complex IV Spur and the Magellanic Stream (MS), respectively. 
The clumps have an angular diameter of  $\lesssim3.5\arcmin$. Assuming that they 
are spherical and located at the same distance of about $1$\,kpc as for IV Spur 
\citep{wakker01} and about $50$\,kpc as for the MS \citep{yoshizawa_noguchi03}, 
the clumps have diameters of about $0.5\ldots0.8$\,pc (QSO\,B1331$+$170) and 
$30\ldots50$\,pc (QSO\,J0003$-$2323), respectively. From this, we 
are able to estimate also other distance-dependent physical parameters such as 
mass and pressure of the gas. The total \ion{H}{i} mass as a function of 
distance $d$ is given by

\begin{equation}
M_\mathrm{HI}=m_\mathrm{H} d^2 \tan^2 \phi \sum_i N_\mathrm{HI}^{(i)}\,,
\label{totalmass}
\end{equation}

with $m_\mathrm{HI}$ being the mass of the hydrogen atom, $\phi$ the angular 
size of a resolution element (pixel) of the $N_\mathrm{HI}$ column density map 
($16\arcsec$ for QSO\,B1331$+$170 and $6\arcsec$ for QSO\,J0003$-$2323) and 
$N_\mathrm{HI}^{(i)}$ the detected column density for each pixel in the map.
The mean density of each clump can be determined using the measured peak column 
density, assuming a spherical symmetry of the clumps and a constant particle 
density $n_\mathrm{HI}$:

\begin{equation}
n_\mathrm{HI}=\frac{N_\mathrm{HI}}{d \tan \phi}\,.
\label{meandensity}
\end{equation}

Using the upper limit for the kinetic temperature, the derived particle density, 
and assuming an ideal gas, one gets an expression for the upper pressure limit

\begin{equation}
\frac{P}{k_\mathrm{B}}=n_\mathrm{HI}T_\mathrm{max}\,.
\label{upperpressurelimit}
\end{equation}

Inserting Equation\,(\ref{meandensity}) into Equation\,(\ref{upperpressurelimit}) 
provides the upper gas pressure limit (under the assumptions made before)
\begin{equation}
\frac{P}{k_\mathrm{B}}=\frac{N_\mathrm{HI}}{d \tan 
\phi}T_\mathrm{max}=\frac{\sin b}{z}\frac{N_\mathrm{HI}}{\tan \phi}T_\mathrm{max}
\label{equpperpressuretodistance}
\end{equation}
as a function of the (unknown) distance of the cloud, or as a function of height, 
$z$, above the plane (considering the Galactic latitude, $b$). However, one 
major uncertainty here is the determination of angular sizes of the clumps, 
which could have a great impact if the clumps are not resolved.

\begin{table}
\caption[Summary of the distance-dependant parameters]{Distance-dependent 
parameters of the small-scale structures observed towards QSO\,B1331$+$170 (WSRT) and QSO\,J0003$-$2323 (VLA).  }
\label{tab_WSRT_clumps_distdep}
\small
\centering
\begin{tabular}{cccccc}\hline\hline
\rule{0pt}{3ex}Quasar&clump&$r$&$M_\mathrm{HI}$&$n_\mathrm{HI}$&$\left(\frac{P}{
k_\mathrm{B}}\right)_\mathrm{upper}$\\
\rule{0pt}{3ex}&&[pc]&[$M_{\sun}$]&[cm$^{-3}$]&[cm$^{-3}$ K]\\\hline
\rule{0pt}{3ex}QSO\,J0003$-$2323
&A&45&470&0.14&27 \\
&B&45&280&0.06&14 \\
&C&49&160&0.06&14 \\
&D&36&150&0.06&45 \\
&E&32&160&0.07&8 \\\hline
\rule{0pt}{3ex}QSO\,B1331$+$170
&A&0.8&0.05&2.5&900 \\
&B&0.6&0.07&5.0&1100 \\
&C&0.5&0.02&5.5&3800 \\
&D&0.6&0.06&5.0&2600 \\
&E&0.5&0.02&4.8&700 \\
&F&0.5&0.05&7.9&600 \\\hline
% \rule{0pt}{3ex}QSO\,B0450$+$1310B
% &A&--&--&--&--&$4.5 \pm 1$ \\
% &B&--&--&--&--&$5 \pm 1.5$ \\\hline
% \rule{0pt}{3ex}J081331$+$254503
% &B&--&--&--&--&$12 \pm 4$ \\
% &C&--&--&--&--&$18 \pm 4$ \\\hline
\end{tabular}
\end{table}

Table\,\ref{tab_WSRT_clumps_distdep} shows the distance-dependent parameters for 
the clumps observed in the directions of QSO\,B1331$+$170 (1\,kpc) and QSO\,J0003$-$2323 (50\,kpc), 
for which a distance was assumed based on their association with known IVC or HVC 
complexes. The columns contain the the spatial diameters, the masses, the particle densities, and the upper pressure limits of the clumps.

%For the determination of the parameters we assume a spherical symmetry for the clumps and a distance of 50\,kpc (MS towards QSO\,J0003$-$2323) and 1\,kpc (IV-spur towards QSO\,B1331$+$170), respectively. 

The small line widths of the clumps, especially the ones towards QSO\,J0003$-
$2323 (Section\,\ref{QSOJ0003} and Table\,\ref{tab_WSRT_clumps}), indicate that 
the absorbing gas is cold with upper temperature limits of $T_\mathrm{kin} 
\approx 100$\,K for the clumps with the most narrow line widths of about $\Delta 
v_\mathrm{FWHM} \approx 2$\,km\,s$^{-1}$ and $T_\mathrm{kin} \approx 3700$\,K 
for the structures with the broadest values of $\Delta v_\mathrm{FWHM} \approx 
13$\,km\,s$^{-1}$. The temperatures could be even lower because effects like 
turbulent motions within the gas likely contribute to the observed line width.
The temperatures and projected sizes of the clumps detected in the direction of 
QSO\,B1331$+$170 and QSO\,J0003$-$2323 show that the clumps are compact and cold 
rather than diffuse and extended.

Due to the fact that the smallest spatial structures observed are of similar 
size as the synthesised WSRT and VLA beams
it is possible that the observed clumps actually are not resolved and contain 
structures on even smaller scales. This raises the question whether IVCs and 
HVCs and their small-scale structures could have fractal properties, as 
discussed in \citet{vogellar_wakker_94}. However, the answer to this question 
requires many more observations at different spatial resolutions.
 
We can now compare the determined upper pressure limits for the high-velocity 
absorbing system in the direction of QSO\,J0003$-$2323 and QSO\,B1331$+$170 with 
the models of \citet{wolfiremckeehollenbachtielens95}. These authors calculated 
the thermal equilibrium gas temperature for HVCs in the Galactic halo as a 
function of height $z$ above the Galactic plane and find that stable two-phase 
gas exists only within a narrow pressure range. Note that both sources lie near 
the Galactic poles, hence the distance roughly equals the Galactic height. 
Various values for metallicities and dust-to-gas ratios were used to take 
different IVC and HVC origins into account. Under the assumption that the observed 
HVC in the direction of QSO\,J0003$-$2323 is associated with the Magellanic 
Stream, we compare the derived upper pressure limit of 
$(P/\mathrm{k}_\mathrm{B})_\mathrm{upper}$ with the pressure range of $P \approx 
20 \ldots 180$\,cm$^{-3}$\,K for the HVC gas stripped from the MS, calculated 
using a metallicity of $Z\approx 0.4$ and a dust-to-gas ratio 
of $D/G=0.2$ \citep{wolfiremckeehollenbachtielens95}. 
To remind the reader, $D/G$ refers to the mass of the small grain population, where $D/G=1$ implies that about 4\% of the cosmic carbon is found in grains smaller than $15\AA{}$.

The observed values of clump A and D are relatively low but consistent with the 
models. The other structures reveal pressures not compatible with the 
simulations. However, as discussed before, the physical size of the clumps might 
be overestimated (if unresolved). In that case, the derived upper pressure limit 
would be underestimated.

For QSO\,B1331$+$170 the simulated pressure range of $P \approx 150 \ldots 
650$\,cm$^{-3}$\,K (assuming a Galactic fountain origin, a metallicity of about 
1.0 solar and a dust-to-gas ratio of $D/G=0.3$) or $P \approx 650 \ldots 
3500$\,cm$^{-3}$\,K (metallicity of about 1.0 and a dust-to-gas ratio of 
$D/G=1.0$) is consistent with our observations. 

For the two other sightlines toward QSO\,B0450$-$1310 and J081331$+$254503, the 
lack of distance information for the halo clouds in these directions 
unfortunately impedes a similar pressure analysis.

\subsection{On the metal abundances of the clumps}

To better understand the nature and the origin of these low-column density, 
small-scale clumps in the halo and their relation to large IVC and HVC complexes 
it would be very helpful to constrain the chemical composition of these objects.
Our optical observations provide column density information for the two species 
\ion{Ca}{ii} and \ion{Na}{i}, while the 21-cm observations give $N$(\ion{H}{i}).

However, there are a number of systematic effects that inhibit a reliable 
estimate for the absolute calcium and sodium abundances in the clouds.  
In diffuse neutral gas, \ion{Ca}{ii} and \ion{Na}{i} are not the dominant 
ionization stages of these elements because of their low ionization potentials 
compared to hydrogen and both elements usually are depleted into dust grains. 
Therefore, the measured \ion{Ca}{ii} and \ion{Na}{i} column densities in the 
halo absorbers are not expected to be representative for the total calcium and 
sodium abundances in the gas. Another severe systematic problem for determining 
the absolute abundances of these elements comes from the comparison of the 
absorption-line data with the \ion{H}{i} 21-cm emission measurements.
The beam sizes of the WSRT and the VLA still are very large compared to the 
pencil beam of the absorption measurements. Therefore, the \ion{H}{i} column 
density values are just averages over a larger area on the sky, and the values 
at the quasar positions are not accurately known. In all our cases the quasar 
sightlines are located outside or at the edge of the \ion{H}{i} clumps. There, 
one might just observe the Gaussian decline of the point spread function in a 
region where there might not even be any detectable signal at all. Also, the 
radial velocities and the component structure of the absorption lines are 
different than those seen in 21-cm emission. This points towards small-scale 
structure in the gas that is not detected and/or resolved in the beam-smeared 
\ion{H}{i} data.   

Despite all these substantial uncertainties, ionic gas phase abundances, $A$, 
for \ion{Ca}{ii} and \ion{Na}{i} (relative to \ion{H}{i}) have been published 
for a large number of Milky Way disk and halo absorbers \cite[e.g.,][]{wakker01, wakkermathis00},
implying that for each ion $A$ scales with \ion{H}{i} 
in a different manner and with different scatter. As an example, we compare the 
ionic gas phase abundance of Na in the halo absorber towards QSO\,B0450$+$1310B 
with these previous observations. We choose the QSO\,B0450$+$1310B sightlines as 
it passes relatively close to one of the \ion{H}{i} clumps seen in the high-
resolution 21-cm data (Fig.\,3). For the halo absorber towards QSO\,B0450$+$1310B 
a column density of $N_\mathrm{HI}=2.8 \cdot 10^{18}$\,cm$^{-2}$  was measured 
by the VLA ($N_\mathrm{Na}=10^{12}$\,cm$^{-2}$). The associated \ion{Na}{i} 
abundance is $A_\mathrm{NaI} \approx 360 \cdot 10^{-9}$, which is about seven 
times larger than \ion{Na}{i} abundances previously measured in diffuse halo 
clouds \citep[$A_\mathrm{NaI} \leq 50 \cdot 10^{-9}$, $A_\mathrm{CaII}=3 \ldots 350 
\cdot 10^{-9}$;][]{wakker01, wakkermathis00}. However, \citet{wakker01} 
points out that the \ion{Ca}{ii} (\ion{Na}{i}) abundance in IVC and HVC gas varies 
by a factor of 2--5 (10) within and by a factor 5--10 (100) between clouds. 
Since all our clouds generally are of low column density, this substantial 
scatter makes it impossible to draw any meaningful conclusions about the Ca and Na 
abundances and overall metallicities of the absorbers based solely on our 
\ion{Ca}{ii} and \ion{Na}{i} measurements.

A much more reliable method to study the metal content of weak halo absorbers is 
the analysis of UV and FUV spectra, as obtained with FUSE and HST/STIS. \citet{richteretal08} 
recently have studied the metallicities in seven low-column 
density high-velocity absorbers by comparing the column densities of \ion{O}{i} 
with that of \ion{H}{i}. The ratio \ion{O}{i}/\ion{H}{i} is a very robust measure for 
the metallicity of the gas, as it does not depend on ionization effects and dust 
depletion. In this study, $N_\ion{H}{i}$ has been derived from FUV absorption 
data by fitting the higher Lyman series lines in absorption against the quasars, 
but not from beam-smeared 21-cm emission observations. The UV measurements 
indicate that similar as for the large IVC and HVC complexes, low-column 
density halo clouds span an abundance range between 0.1 and 1.0 solar, 
suggesting multiple (galactic and extragalactic) origins for the absorbers. Also 
relative abundance ratios (e.g., Fe/Si) have been used to constrain the 
enrichment history of these clouds \citep[see][]{richteretal08}. 
Interestingly, the simultaneous detection of \ion{Ca}{ii} and other low ions 
such as \ion{O}{i}, \ion{Si}{ii}, and \ion{Fe}{ii} in two high-velocity systems 
towards PHL\,1811 implies that the high-velocity \ion{Ca}{ii} absorbers 
considered in this study and the weak UV absorption systems studied by \citet{richteretal08} trace the same 
population of low-column density halo clouds. 

\section{Conclusions}\label{secsummary}

In this paper we have presented high-resolution \ion{H}{i} observations with the 
WSRT and the VLA for four lines of sight through the Galactic halo were we previously detected weak low-, intermediate-, and high-velocity gas in \ion{Ca}{ii} and \ion{Na}{i} absorption.
Along these four sightlines, we detected one high-velocity cloud (probably belonging to the Magellanic Stream) seen in \ion{Ca}{ii} 
and \ion{Na}{i} absorption and 21-cm emission. The other four detected features show low-velocity gas ($v_\mathrm{dev}<30\,\mathrm{km\,s}^{-1}$). One of these systems is likely associated with the IV spur, another one is eventually part of the LLIV arch, both being IVC complexes. For the two remaining clouds we can not finally decide whether they are part of the Milky Way disk or are located in the (lower) Galactic halo. The major results of our study are:

\begin{enumerate}

\item Along all four sightlines our high-resolution \ion{H}{i} observations 
reveal the presence of cold, compact (sub-pc scale, except those associated with the 
Magellanic Stream) neutral gas clumps associated with 
\ion{Ca}{ii} and \ion{Na}{i} absorbers. The intermediate- and high-velocity systems are located in the 
Milky Way halo, whereas the low-velocity features are either located in the lower halo or in the Milky Way disk.

\item The detected clumps have low peak column densities of $N_\mathrm{HI}\leq3 
\cdot 10^{19}$\,cm$^{-2}$. This, together with their small angular sizes ($1-3$\,arcmin, typically), 
suggest that these small-scale \ion{H}{i} structures  are unseen and/or unresolved in current \ion{H}{i} 21-cm all-sky surveys.

\item The \ion{H}{i} line widths of the clumpy structures are relatively narrow 
($1.8 \leq \Delta v_\mathrm{FWHM} \leq 13$\,km\,s$^{-1}$), leading to upper 
limits for the kinetic temperature in the range $T_\mathrm{kin}^\mathrm{max}=70-
3700$\,K.

\item In all four directions, the quasar sightlines used for the absorption 
spectra pass the outer envelopes of the clumps or the inter-clump medium, but do 
not pass the clump-cores. The presence of narrow \ion{Ca}{ii} and \ion{Na}{i} 
absorption along these sightlines demonstrates that additional small-scale 
structure exists at column densities below the detection limit of the high-
resolution \ion{H}{i} 21-cm observations.    

\item The sky positions as well as the radial velocities of the halo absorber 
towards QSO\,J0003$-$2323 and QSO\,B1331$+$170 indicate a possible association 
with the Magellanic Stream and IV Spur, respectively. Using available distance 
information for these halo clouds we are able constrain distance-dependent 
physical parameters such as gas mass and thermal pressure. The derived 
parameters match the values predicted by the models of Wolfire et al.\,(1995).

\end{enumerate}

Our study indicates that there exists a substantial degree of small-scale 
structure in the neutral galactic halo gas that can only be traced with high-
resolution \ion{H}{i} 21-cm observations and with quasar absorption-line 
spectroscopy. The combination of the various data sets shows, how important 
measurements at different wavelengths and resolutions are to get a more 
realistic view of the distribution and physical properties of neutral gas 
structures in the Milky Way halo. 

Using radio synthesis telescopes, we will continue the search for small-scale 
structures that are not resolved with the Effelsberg telescope. Moreover, we 
will proceed with our effort to systematically study absorption features in the 
Galactic halo along quasar sightlines using optical instruments and future 
space-based UV spectrographs. Such combined high-resolution emission and 
absorption observations will be of crucial importance to better understand the 
spatial distribution of gaseous structure in the inner and outer halo of the 
Milky Way and their relation to intervening quasar absorption-line systems at 
low and high redshift.

\begin{acknowledgements} 

P.R. and N.B.B. acknowledge financial support by the German
\emph{Deut\-sche For\-schungs\-ge\-mein\-schaft}, DFG,
through Emmy-Noether grant Ri 1124/3-1.

\end{acknowledgements}

\bibliographystyle{aa}

\bibliography{references}

\end{document}